\documentclass[12pt]{iopart}

\usepackage{amsfonts}
\expandafter\let\csname equation*\endcsname\relax
\expandafter\let\csname endequation*\endcsname\relax
\usepackage{amsmath}
\usepackage{amssymb}
\usepackage[square, numbers, sort&compress, nonamebreak]{natbib}
\usepackage{graphicx}
\usepackage{color}\usepackage{url}
\usepackage[bookmarks, breaklinks]{hyperref}

\begin{document}
 \title{Brillouin zone labelling for quasicrystals}
\author{Jean-Marc Gambaudo and Patrizia Vignolo}
\address{Universit\'e de Nice - Sophia Antipolis, Institut non Lin\'eaire de Nice, CNRS, 1361 route des Lucioles, 06560 Valbonne, France}
\ead{jean-marc.gambaudo@inln.cnrs.fr; patrizia.vignolo@inln.cnrs.fr}

\begin{abstract}
We propose a scheme to determine the energy-band dispersion 
of quasicrystals which does not require any periodic approximation
and which directly provides the correct structure of the extended Brillouin 
zones. 
In the gap labelling viewpoint, this allow to transpose the measure
of the integrated density-of-states to the measure of the effective 
Brillouin-zone areas
that are uniquely determined by the position of the Bragg peaks. Moreover 
we show that the Bragg vectors can be determined by the stability analysis 
of the law of recurrence used to generate the  quasicrystal. 
Our analysis of the gap labelling in the quasi-momentum space
opens the way to an experimental proof of the gap labelling itself within the framework of an optics experiment, polaritons, 
or with ultracold atoms.
\end{abstract}

\section*{Introduction}
Quasicrystals are alloys that show long range order but possess symmetries that prevent them to be crystals. The first examples observed were rapidly quenched alloys of Aluminum and Manganese exhibiting icosahedral symmetry.
From their spectacular experimental realization in the early 80's \cite{Shechman1984} to their very recent discovery as natural objects in the Kamtchatka mountains  \cite{Bindi2009}, quasicrystals have been the subject of very active research, whose domains extend far beyond the scope of solid state physics. 
Quasicrystals can also 
be  reproduced with dielectric
materials \cite{Man2005} or via the interference of laser beams 
\cite{Guidoni1999} and thus they are 
studied not only for the transport of electrons, but also of 
light, matter waves or even 
quasiparticles such as polaritons \cite{Bloch-fibo}.
For all these cases the knowledge of the 
energy-band dispersion is  a crucial concept which is much more  sophisticated than in the crystal case.

For a periodic crystal, the diffraction spectrum is pure point with a 
lattice structure and the 
energy spectrum is organized in bands separated by gaps.
All bands can be calculated in the fundamental Brillouin zone (BZ) that
is the primitive cell of the reciprocal lattice. This is possible 
because of the equivalence of the BZs, meaning that the area of 
each zone is the same and that different BZs 
 are connected by the Bragg vectors. It follows that the number 
of states in each band is the same and that the integrated density of states
(IDOS) ${\mathcal N}(E)$ is a staircase with identical steps.
For quasicrystals the situation is a bit different. The diffraction spectrum is again pure point but it does not exist a 
fundamental BZ. On the other hand, the IDOS is again a staircase, despite its more sophisticated structure, and thus one can continue to speak about 
bands and gaps. The heights of the steps are all different and one 
can use these heights to label the gaps. One can enumerate the gaps in a way that depends only on the geometry of the quasicrystal and not on the nature of the potential it induces. In particular, for a potential which varies with a parameter, the labelling remains invariant under the parameter evolution (this is the subject of the so-called gap labelling theory~\cite{Bellissard1993,Benameur2003,Kaminker2003,Bellissard2006}). For instance it was shown that for Fibonacci or Penrose tilings, in the limit of an infinite sample of quasicrystal, and when $E$ belongs to a gap, 
the IDOS reads: 
\begin{equation}
{\mathcal N}(E)={z}_{i_1}+\lambda{z}_{i_2}
\end{equation}
where $\lambda=(\sqrt{5}+1)/2$ is the golden number and the $z_{i}$'s are  relative integers.
Morover, the staircase is a devil's staircase:  each step is divided in smaller steps and each smaller step is divided
in steps even smaller, etc\dots, and the spectrum is a Cantor set. 
At a finite but large fixed energy resolution one can identify
the main gaps of the staircase and get in theory a good measure of  
${\mathcal N}(E)$ when $E$ belongs to a gap, {\it i.e.} the gap labelling.
With conventional quasicrystal alloys,  
the measure of ${\mathcal N}(E)$ is completely inaccessible in the Fermi sea.
However with matter waves
it could be measured in some particular cases such as spin-polarized fermions or Tonks bosons \cite{Lang2012,Cai2011}.

In this paper we introduce a very efficient method that allow us to 
calculate the band-energy dispersion
for quasicrystals as a function of the (effective) extended BZs
and we apply it for the case of a Penrose-tiled quasicrystal.
This explicit 
relation between the spectrum and the
quasi-momenta allows to pass from the gap labelling to 
the {\it Brillouin zone labelling}. 
The importance of this step is both conceptual and pratical since 
the extended BZs are more easily accessible than the IDOS and
can be measured in optics \cite{Man2005}, with polaritons \cite{Bloch-honey} 
or using matter waves \cite{Greiner2001,Wirth2011}.

The paper is organized as following. In Sec. \ref{sec-system} we briefly
introduce the Penrose-tiled quasicrystal that 
we have choosen as specific example for our analysis. In Sec. \ref{sec-band}
we illustrate the scheme that we have developed to determine 
the energy-band dispersion for 
quasicrystals as a function of the extended BZs. 
Using the band structure calculation as starting point, in this section 
we discuss the
reintepretation of the gap labelling in terms of a BZ labelling.
On the other hand the BZ labelling can be deduced by a Bragg peak labelling.
This is shown in Sec. \ref{sec-bragg} where we demonstrate that it 
is possible to determine the Bragg vectors, and select the brightest ones, 
by the stability analysis 
of the law of recurrence used to generate the  quasicrystal. 
Our concluding remarks on the possibility to access experimentally
to the BZ labelling, and thus to the gap labelling, 
are given in Sec. \ref{sec-concl}.

\section{The system}
\label{sec-system}
\begin{figure}
\begin{center}
\includegraphics[width=0.6\linewidth]{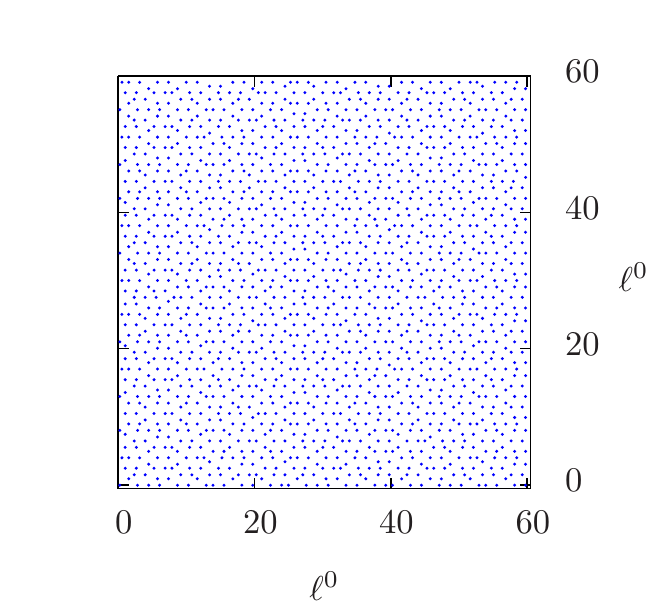}
\caption{\label{fig-punti}Positions of the vertices of the quasicrystal tiling
considered in this work, in unit of $\ell^0$, $\ell^0$ being the 
minimum distance between 2 points of the tiling.}
\end{center}

\end{figure}

\begin{figure}
\begin{center}
\includegraphics[width=0.6\linewidth]{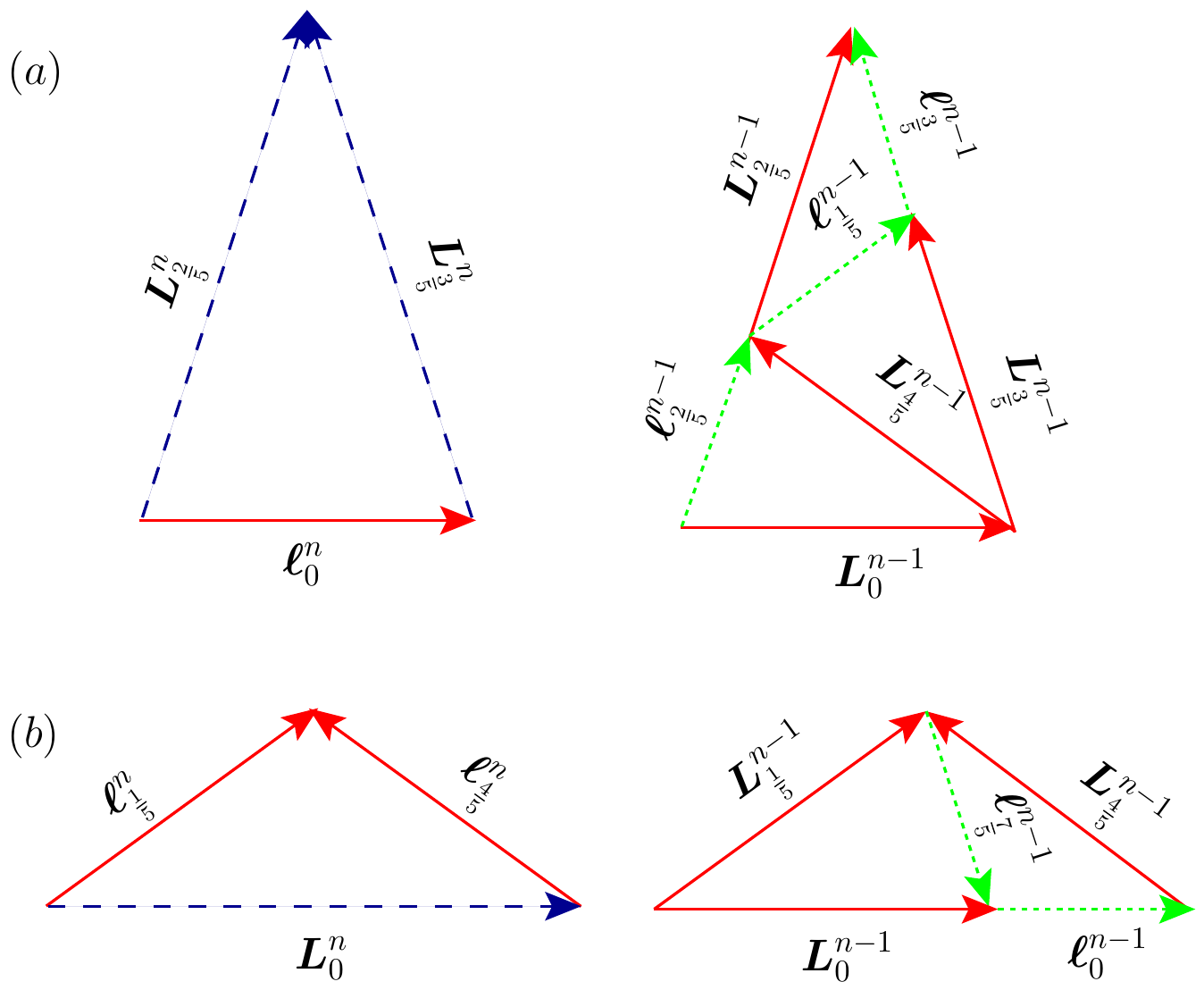}
\caption{\label{fig-sub}Robinson triangle decomposition for a Penrose tiling.
(a) tiling of $\Delta_1^n$ and (b) of $\Delta_2^n$ 
by $\Delta_1^{n-1}$ and $\Delta_2^{n-1}$.}
\end{center}

\end{figure}

We consider the two-dimensional (2D) Hamiltonian
\begin{equation}
H=-\dfrac{\hbar^2\nabla^2}{2m}+V(\mathbf{r})
\end{equation}
for a particle of mass $m$,
where $V(\mathbf r)=V_0\delta(\mathbf{r}-\mathbf{r}_i)$,
$\mathbf{r}_i$ being the positions of the vertices of a quasicrystal
tiling (see Fig. \ref{fig-punti}). 
In this paper we focus on the specific case of
a Penrose-tiled quasicrystal obtained using the Robinson triangle 
decomposition shown in Fig. \ref{fig-sub}.
The tiling is composed with two types of tiles: the first is the triangle 
$\Delta_1^0$ with edges $\{\ell^0,L^0,L^0\}$, the second the triangle 
$\Delta_2^0$ 
with edges $\{L^0,\ell^0,\ell^0\}$ , with $L^0/\ell^0=\lambda$.
The triangles $\Delta_j^{0}$ (with $j=1$ or 2) are embedded in the triangles
$\Delta_j^{1}$ that are embedded  in $\Delta_j^{2}$ etc\dots, the lengths of the
edges satisfying  the recursive relations 
$\ell^{n}=L^{n-1}$ and $L^{n}=\ell^{n-1}+L^{n-1}$ (see Fig. \ref{fig-sub}).
The internal angles are equal to $\{\pi/5,2\pi/5,2\pi/5\}$ 
for all the triangles $\Delta_1^{n}$, and to 
$\{3\pi/5,\pi/5,\pi/5\}$ for all the triangles $\Delta_2^{n}$.
By construction the Penrose-tiles quasicrystal is self-similar, indeed
the same patterns, $\Delta_1^{n}$ and $\Delta_2^{n}$, 
occur at different spatial scales, but it lacks any translational symmetry
(it is not a periodic crystal).

Potentials with the same symmetry as the Penrose-tiled quasicrystal 
can be realized in optics and microwaves with 
dielectric materials \cite{Man2005,Bellec2013} or arrays 
of waveguides \cite{Rechtsman2013}, 
in polariton experiments by a 
suitable engineering of 
the photonic component \cite{Tanese2013,Bloch-fibo,Bloch-honey},
in ultracold atoms via the interference of five beams \cite{Guidoni1999} or 
by using a Spatial Light Modulator (SLM) device \cite{Barboza2013}. 
\section{Energy-band dispersion for quasicrystals}
\label{sec-band}
Let us  consider the subspace $\mathcal{O}_{\mathbf{k}}$ of the plane waves 
$W_{j,\mathbf{k}}(\mathbf{r})=e^{-i(\mathbf{k}+\tilde{\mathbf{q}}_j)\cdot\mathbf{r}}/\sqrt{S}$, $S$
being the quasicrystal surface area. The vectors $\tilde{\mathbf{q}}_j$ are  generated by linear combinations of the Bragg vectors $\mathbf{q}_j$ 
with relative integer coefficients, namely $\tilde{\mathbf{q}}_j=\sum_i z_i\mathbf{q}_i$. The $\mathbf{q}_j$ satisfy
$\hat V(\mathbf{q}_j)=\int e^{-i{\mathbf q}_j\cdot{\mathbf r}}
V(\mathbf r)d\mathbf r\neq 0$.
Analogously to the periodic case, the subspace $\mathcal{O}_{\mathbf{k}}$ 
is closed under the application of the operator $H$. Thus the diagonalization 
of 
\begin{equation}
\begin{split}
H_{\mathbf{k}}&=\int W_{j,\mathbf{k}}(\mathbf{r})^*H\,W_{i,\mathbf{k}}(\mathbf{r})\,d\mathbf{r}\\
&=\dfrac{\hbar^2|\mathbf{k}+\tilde{\mathbf{q}}_j|^2}{2m}\delta_{i,j}+\hat
V(\tilde{\mathbf{q}}_i-\tilde{\mathbf{q}}_j),
\end{split}
\end{equation}
 within the subspace $\mathcal{O}_{\mathbf{k}}$, provides rigorous 
eigenfunctions and eigenvalues of $H$ \cite{Kaliteevski2000}. The problem is 
how to associate
to each eigenvalue its corresponding $\mathbf{k}$ vector.
Usually a periodic approximation is done by considering 
the first effective BZ  
as adequate to describe all bands \cite{Kaliteevski2000,Florescu2009}. 
But this is a rather rough approximation 
since, as we will show below, the BZs
have  different areas: for instance, the ratio between the 
area ${\mathcal S}_2$ of the second zone and area ${\mathcal S}_1$ 
of the first one is ${\mathcal S}_2/{\mathcal S}_1=\lambda-1$. 
 
Our strategy to determine the energy-band dispersion,
is the following. First we make a selection 
of the $\tilde{\mathbf{q}}_j$'s vectors, by fixing a threshold for the value of 
$|\hat V({\mathbf{q}}_j)|^2$ and a maximum value of $\tilde{q}_j^2$ (this value depends on the maximum value of $k^2$ we are interested in). In this way we obtain a truncated finite subspace $\tilde{\mathcal{O}}_{\mathbf{k}}$ of dimension $\tilde N$, $\tilde N$ being the number of selected $\tilde{\mathbf{q}}_j$, and
we fix the level in the hierarchy we are able to observe the gaps.
Then we diagonalized $H$ whithin the subspace $\tilde{\mathcal O}_{\mathbf{k}}$ getting $\tilde N$ eigenvalues.
With the aim to determine the dispersion relation $E(\mathbf{k})$, for each vector $\mathbf{k}$ we select the ``physical'' eigenvalue $E_{\bar{i}}$, among the $\tilde N$ ones, that satisfies the condition $\lim_{V_0\rightarrow 0} E_{\bar{i}}\rightarrow (\hbar k)^2/2m$. From an operational point of view, 
we sort the $\tilde N$ 
eigenvalues by increasing values 
\begin{equation}
E_1<E_2<\dots E_{\tilde N}
\end{equation}
 and similarly we order the quantities $K_j=|{\mathbf{k}}+\tilde{\mathbf{q}}_j|^2$, 
\begin{equation}
[K_j]_1<[K_j]_2<\dots[K_j]_{\tilde{N}}.
\end{equation} 
We select the eigenvalue $E_{\bar{i}}$ if $[K_j]_{\bar{i}}=k^2$.
The points of exact degeneracy $[K_j]_{\alpha}=[K_j]_{\beta}$ are not taken into account by varing the vector ${\mathbf{k}}$ uniformly, but avoiding the points ${\mathbf{k}}=-\tilde{\mathbf{q}}_j$.

This procedure allows to assign each eigenvalue to the corresponding 
quasi-momentum and thus it defines the band structure and the extended BZ 
unequivocably, if they exist, without requiring the existence 
of a primitive cell of the reciprocal lattice, namely 
the system to be periodic.
The variant with respect to the periodic case is that, since it does 
not exist a fundamental BZ, we have to repeat the procedure 
for any $\mathbf k$ vector we are interested in. 

At fixed ${\mathbf{k}}$, the unselected $\tilde N-1$ eigenvalues correspond
to eigenvalues belonging to other Brillouin zones as in the periodic case.
However, away from the center of the first Brillouin zone, the lack of periodicity in the $\tilde{\mathbf{q}}_j$ space, makes non-trivial the allocation of these eigenvalues with respect to their corresponding Brillouin zone.

\begin{figure}
\begin{center}
\includegraphics[width=0.56\linewidth]{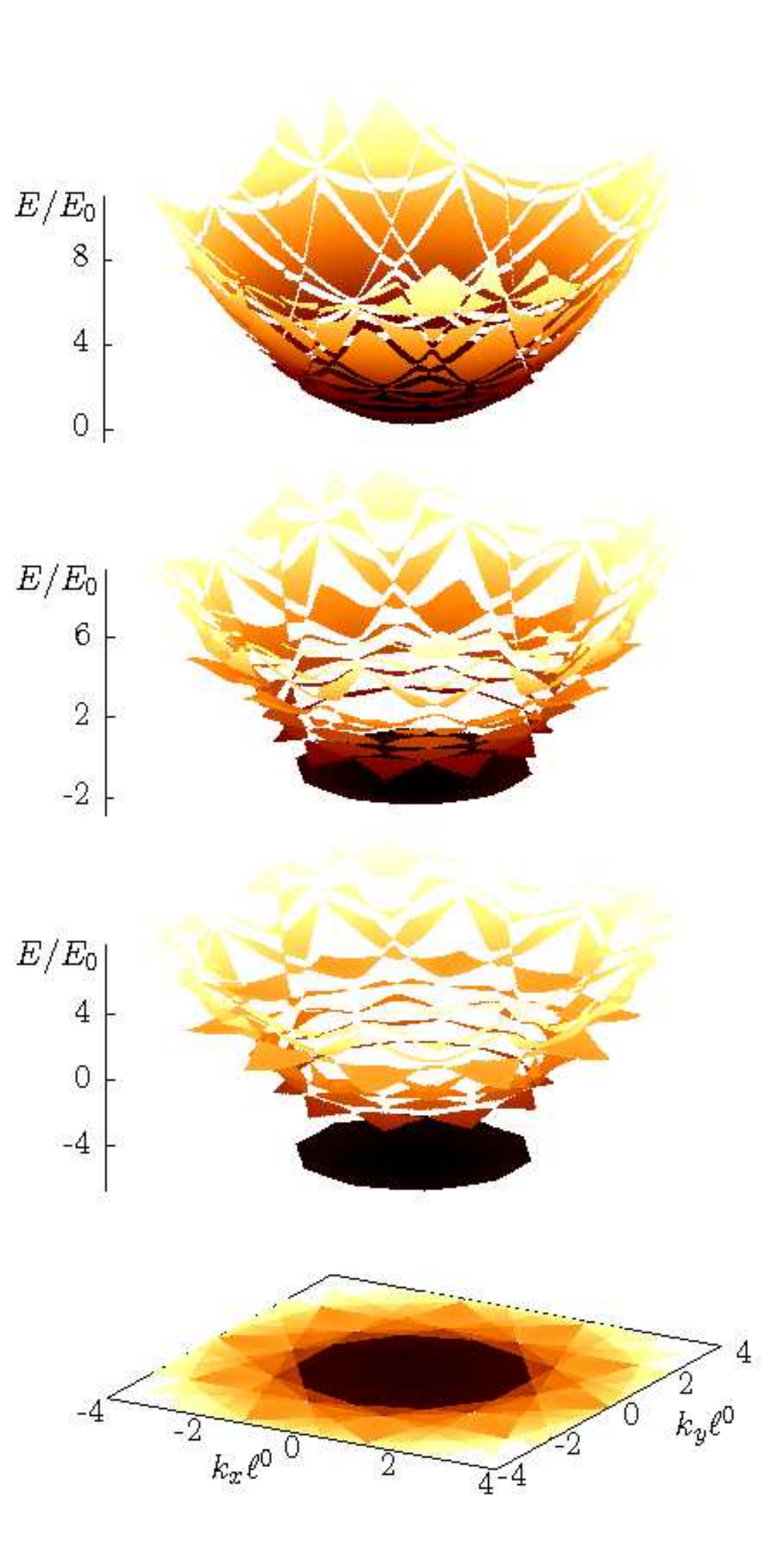}
\caption{\label{fig-bande}Energy-band dispersion $E({\mathbf{k}})$, in unit of $E_0=(\hbar/\ell^0)^2/m$ as a function of $k_x$ and $k_y$, in unit of $1/\ell^0$, for a Penrose-tiled potential. From top to bottom: $V_0=-1$, -3, -5 $E_0$.
The color contrast in the map view shows the extended Brillouin zones.
These curves have been obtained by using $\tilde N=31$ and the selected vectors $\mathbf{q}_i$ correspond to the brightest Bragg spots (see Fig. \ref{fig-bragg}), by fixing a threshold for the value of $|\hat V({\mathbf{q}}_j)|^2$ at 0.03 $V_0^2$ and $|{\mathbf{q}}_j|<8/\ell^0$.
We have verified that if we include the linear combination of these vectors in the $\tilde{\mathcal{O}}_{\mathbf{k}}$ subspace, the calculation of the 
spectrum does not 
change significantly. The inclusion of $\mathbf{q}_i$'s vectors corresponding to less bright Bragg spots would introduce a further fragmentation of the spectrum, which would be visible with a higher energy resolution.}
\end{center}
\end{figure}
The energy-band dispersion for our Penrose-tiled potential is shown in 
Fig. \ref{fig-bande}. For a small potential depth (top panel in Fig. 
\ref{fig-bande}), the 
dispersion law is approximately parabolic as for a free-particle,
but one can already observe the opening of the gaps
and the fact that the sizes of the bands, and their projections
on the ${\mathbf k}$-plane (map view in Fig. \ref{fig-bande}), are different 
from each other. 
By increasing the potential depth (central and bottom panels in Fig. 
\ref{fig-bande}) the gaps open further and the bands flatten, but obviously
the surface area of their projections on the ${\mathbf k}$-plane is independent
from the potential strength. These surfaces represent the extended 
BZs. They look like a sunflower whose center 
(the first BZ) is a decagon surrounded by petals of 
different sizes (the higher-order BZs). Remark that the energy-band dispersion
can be measured quite straightforwardly with polaritons \cite{Bloch-fibo,Bloch-honey}.

As expected from the gap labelling theory, the number of states 
${\mathcal N}_i$ in each band
is different from one band to another. This property is intimately 
connected to the splitting of the BZs,
more precisely, in the limit of an infinite sample, the IDOS 
${\mathcal N}_i$ of 
the $i$-th band is proportional
to the area
${\mathcal{S}}_i$ of the $i$-th BZ \cite{Luck1989}. 
This is due to the fact that
${\mathcal N}_i$ depend only on the geometry 
of the crystal and not on the strength of the potential: once the free-particle parabola is fragmented, the number of states per band is incompressible. 
Thus if one consider the case of a 
vanishing potential where the density of states is a constant $\rho(E)\sim E^{(D-2)/2}$, $D$ being the dimensionality of the system, one can easily
deduce that 
\begin{equation}
{\mathcal N}_i =\int_{i-th\,{\rm band}}\!\!\!\rho(E)\,dE\propto\int_{i-th\,{\rm BZ}}\!\!\!dp^2= {\mathcal{S}}_i.
\end{equation} 
Thus the BZs of quasicrystals can be labelled by their
areas and,  for a Penrose-tiled potential, there exist 
two integers $z_{i_1}$ and  $z_{i_2}$ such that:
\begin{equation}
{\mathcal S}_i=s_0(z_{i_1}\lambda+z_{i_2}).
\end{equation} 

\section{Bragg peak labelling}
\label{sec-bragg}
\begin{figure}
\begin{center}
\includegraphics[width=0.6\linewidth]{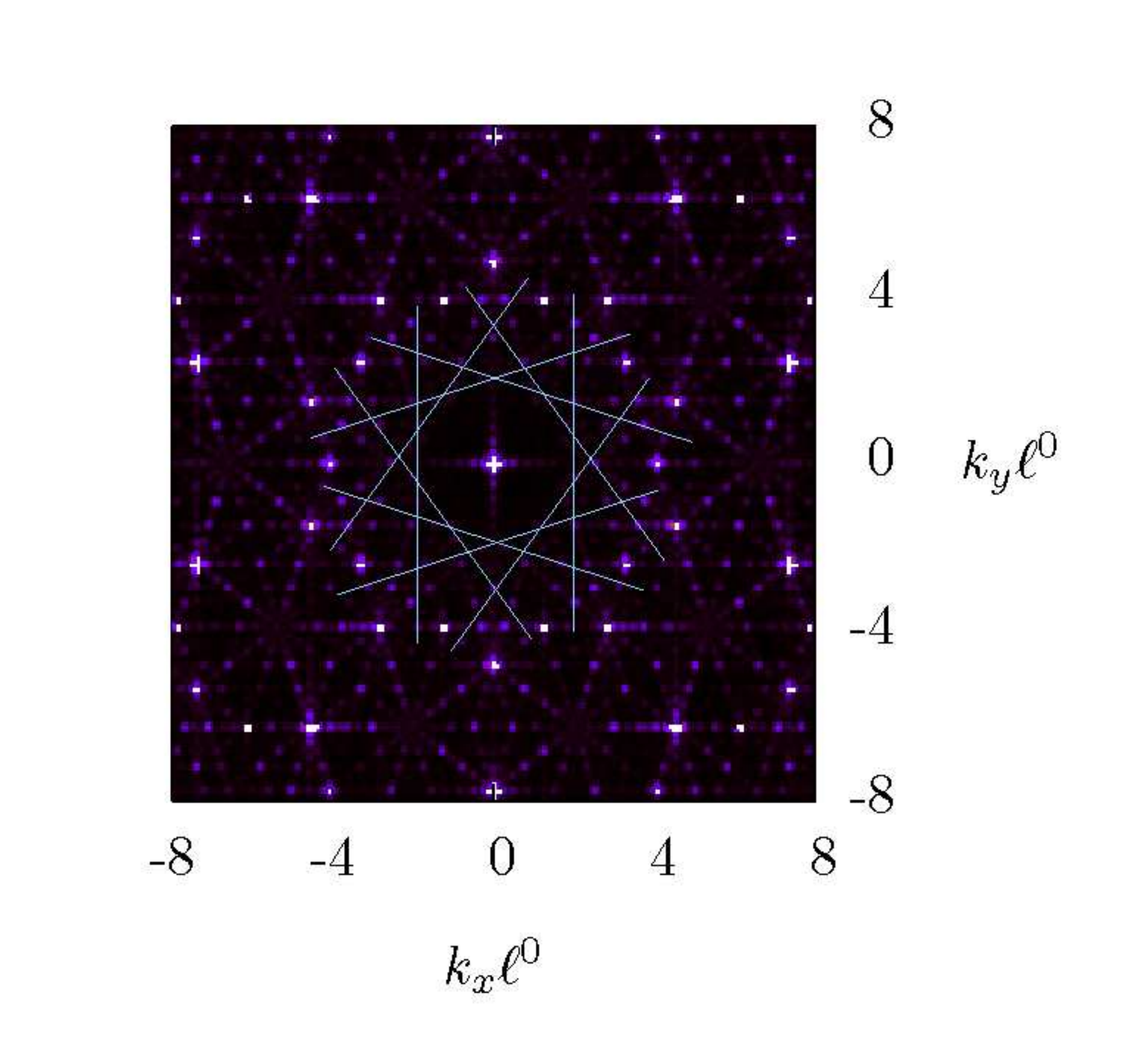}
\caption{\label{fig-bragg} Bragg diffraction of a Penrose-tiled potential 
formed by 1700 potential wells distributed
on a square of side $60\ell^0$ (see Fig.~\ref{fig-punti}). 
The lines are the orthogonal bi-sectors 
of the brightest ${\mathbf{k}}$ Bragg vectors and show the boundaries of the 
first BZs.}
\end{center}
\end{figure}

The knowledge of the energy-band dispersion is not necessary to determine the BZ structure. Making a perturbative reasoning, the gaps open if two free-particle states $|{\mathbf k}_i\rangle$ and  $|{\mathbf k}_j\rangle$,
with the same kinetic energy, ${k}_j^2-k_i^2=0$, are coupled by the potential $V$, namely if 
$\hat V({\mathbf{k}}_j-{\mathbf{k}}_i)\neq 0$, thus if $({\mathbf{k}}_j-{\mathbf{k}}_i)$ is a 
Bragg vector ${\mathbf{q}}_l$ \cite{Grosso}. This leads to
\begin{equation}
({\mathbf{k}}_j-{\mathbf{k}}_i)\cdot({\mathbf{k}}_j+{\mathbf{k}}_i)={\mathbf{q}}_l\cdot(2{\mathbf{k}}_i+{\mathbf{q}}_l)=0,
\end{equation}
namely the boundaries of the BZs are given by the orthogonal bi-sectors of the Bragg vectors $\mathbf{q}_l$ as schematically illustrated 
in Fig. \ref{fig-bragg}.
Therefore the BZ labelling, and thus the gap labelling, can be indirectly deduced by a Bragg diffraction experiments. The more brillant the peak is,  
the more the gap will open effectively as a function of the potential strength. Thus by setting a threshold for the brightness of the peaks, 
we fix the level of the observable gaps, and this is exactly what we do when we select the vectors 
$\tilde{\mathbf{q}}_l$ for the
calculation of the energy-band dispersion (see details of the procedure in the caption of Fig. \ref{fig-bande}).

The ensemble of the Bragg peaks at a fixed threshold for the brightness 
is relatively dense\footnote{A subset $X$ of the plane is relative dense if the distance of any point of the plane to $X$ is uniformely bounded.} and the lower this threshold is, the more the number 
of peaks increases \cite{Elser1985,Kaliteevski2000}. This does not happen with periodic crystals where 
the number and the structure of the Bragg peaks is independent from 
their brightness.
This intriguing phenomenon can be understood 
by the stability analysis of 
the law of recurrence used to generate the Penrose-tiled quasicrystal. 

Let us consider again the Robinson triangle decomposition shown in Fig. 
\ref{fig-sub}.
We introduce the vectorial notation 
\begin{equation}
\boldsymbol{\ell}^n_\alpha =
\ell^n[\cos(\alpha\pi)\,\hat e_x+\sin(\alpha\pi)\,\hat e_y],
\end{equation}
where $\alpha \in \{0, 1/5, 2/5,\dots, 9/5\}$, 
and analogously for $\boldsymbol{L}^n_\alpha$.  For any, $n\geq 0$, it can 
be shown  that the space $\mathcal{A}$ (${\mathbb Z}$-module) of all linear combination with integer coefficients of all the $ \boldsymbol{\ell}^n_\alpha$'s and  the  $\boldsymbol{L}^n_\alpha$ is a 4-dimensional ${\mathbb Z}$-module and the vectors $\boldsymbol{\ell}^n_0, \boldsymbol{\ell}^n_{1/5}, \boldsymbol{\ell}^n_{2/5},  \boldsymbol{\ell}^n_{3/5}, $ form a basis of this module. In this basis, it is easy to check that matrix giving recurrence law for 
the generation of the Penrose tiling ({\it i.e.} the matrix that gives the expression of $\boldsymbol{\ell}^{n+1}_0, \boldsymbol{\ell}^{n+1}_{1/5}, \boldsymbol{\ell}^{n+1}_{2/5},  \boldsymbol{\ell}^{n+1}_{3/5},  $ as linear combination with integer coefficient of the vectors $\boldsymbol{\ell}^{n}_0, \boldsymbol{\ell}^{n}_{1/5}, \boldsymbol{\ell}^{n}_{2/5},  \boldsymbol{\ell}^{n}_{3/5}$)  is given by

\begin{equation}
\mathcal{M}=\left(\begin{matrix}
1&0&1&-1\\
1&0&1&0\\
0&1&0&1\\
-1&1&0&1\\
\end{matrix}
\right).
\end{equation}
The eigenvectors 
of $\mathcal{M}$
are ${\mathbf v}_{+1,+2}^n$ with components $(-\lambda,-1,0,1)$ and $(\lambda,\lambda,1,0)$ and eigenvalues ${m}_{+1}=m_{+2}=\lambda$, and 
 ${\mathbf v}_{-1,-2}^n$ with components $(\lambda-1,-1,0,1)$ and $(1-\lambda,1-\lambda,1,0)$ and 
eigenvalues ${m}_{-1}=m_{-2}=1-\lambda$.
Thus, under iteration,  vectors that belong to the stable eigenspace 
generated by ${\mathbf v}_{-1}^n$ and ${\mathbf v}_{-2}^n$ converge to zero 
exponentially fast.

The diffraction pattern $I(\mathbf r)$ of $V(\mathbf r)$ shown in Fig. \ref{fig-bragg} is given by
\begin{equation}
I(\mathbf r)\propto \left|\hat V(\mathbf k)\right|^2=\left|\int e^{-i{\mathbf k}\cdot{\mathbf r}}
V(\mathbf r)d\mathbf r\right|^2=V_0^2\left|\sum_{j} e^{-i{\mathbf k}\cdot{\mathbf r}_j}\right|^2.
\end{equation}
It is known that for the  Penrose-tiled quasicrystal (or more generally for a larger class of  substitution  tilings) the Bragg peaks correspond to the values of ${\mathbf{k}}$ such that for any  $\alpha \in \{0, 1/5, 2/5,\dots, 9/5\}$, the quantities $e^{-i{\mathbf k}\cdot{\boldsymbol{L}^n_\alpha}}$ and $e^{-i{\mathbf k}\cdot{\boldsymbol{\ell}^n_\alpha}}$ converge to 1 exponentially fast as $n$ goes to $+\infty$ (see \cite{Dworkin1993,Lee2002,Solomyak2005,Barge2006}). Since the vectors $\boldsymbol{\ell}^n_0, \boldsymbol{\ell}^n_{1/5}, \boldsymbol{\ell}^n_{2/5},  \boldsymbol{\ell}^n_{3/5}, $ form a basis of $\mathcal A$, it is enough to show the convergence of  $e^{-i{\mathbf k}\cdot{\boldsymbol{\ell}^n_\alpha}}$ for $\alpha =0, 1/5, 2/5, 3/5$. Notice that, for $\alpha = (j-1)/5$,  
\begin{equation}
\begin{split}
{\mathbf k}\cdot{\boldsymbol{\ell}^n_\alpha} =  {\mathbf k}\cdot { \sum_{j =1}^{j = 4} m^n_{i, j}  {\boldsymbol{\ell}^0_{(j-1)/5}}  =  { \sum_{j =1}^{j =4} m^n_{i, j} { \mathbf k}\cdot {\boldsymbol{\ell}^0_{(j-1)/5}}}}
\end{split}
\end{equation}
 where $m^n_{i, j}$ is the coefficient at the $i^{th}$ line and $j^{th}$ column of the matrix $\mathcal M$.

\begin{figure}
\begin{center}
\includegraphics[width=0.6\linewidth]{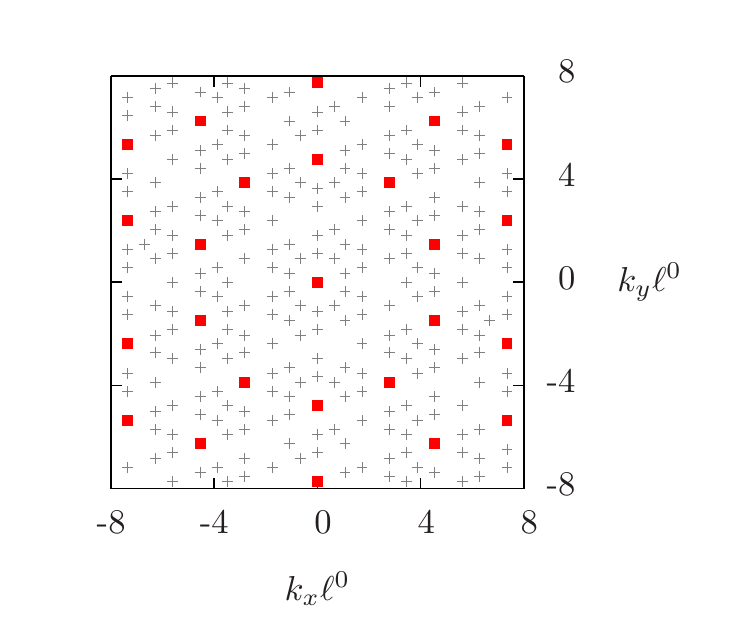}
\caption{\label{fig-picchi} Bragg peaks for a Penrose-tiled quasicrystal
calculated by using Eqs. (\ref{bellissime}). The gray cosses correspond
to $||\alpha{\mathbf v}_{-1}^0+\beta{\mathbf v}_{-2}^0||<20$, while the
red squares correspond to 
$||\alpha{\mathbf v}_{-1}^0+\beta{\mathbf v}_{-2}^0||<5$.}
\end{center}
\end{figure}

It follows from the above discussion that the condition for ${\mathbf k}$ to be a Bragg peak is that
 the vector
$
{\boldsymbol \xi}^n=(\mathbf k\cdot\boldsymbol{\ell}^n_0,\mathbf k\cdot \boldsymbol{\ell}^n_{\frac{1}{5}},\mathbf k\cdot\boldsymbol{\ell}^n_{\frac{2}{5}},\mathbf k\cdot \boldsymbol{\ell}^n_{\frac{3}{5}})$ belongs to the stable eigenspace of the matrix $\mathcal M$ modulo a translation on the lattice $2\pi{\mathbb Z}^4$. That is 

\begin{equation}
{\boldsymbol \xi}^0+(2p_1\pi,2p_2\pi,2p_3\pi,2p_4\pi)=
\alpha{\mathbf v}_{-1}^0+\beta {\mathbf v}_{-2}^0 ,
\end{equation}
for $p_i$'s in ${\mathbb Z}$.
This condition is fulfilled if
\begin{equation}
\begin{split}
&k_x\ell^0=\dfrac{2\pi}{3-\lambda}[p_1-p_3+p_4+\lambda(p_3-p_4)]\\
&k_y\ell^0=\dfrac{2\pi[-p_1+2p_2+p_4-p_3+\lambda(p_4+p_3)]}{2[\sin(3 \pi/5)+2\sin(\pi/5)]}.\\
\label{bellissime}
\end{split}
\end{equation}
The brighest peaks visible in Fig. \ref{fig-bragg} correspond to the
smallest values of $||\alpha{\mathbf v}_{-1}^0+\beta{\mathbf v}_{-2}^0||$;
for these values, the  convergence to 1 of 
the terms $e^{-i{\mathbf k}\cdot{\boldsymbol{L}^n_\alpha}}$ and $e^{-i{\mathbf k}\cdot{\boldsymbol{\ell}^n_\alpha}}$ is achieved more rapidly.
In Fig. \ref{fig-picchi} we show some Bragg peak positions as predicted by 
Eqs. (\ref{bellissime}). We selected them by choosing a maximum threshold
for the value of $||\alpha{\mathbf v}_{-1}^0+\beta{\mathbf v}_{-2}^0||$: 
20 for the gray crosses and 5 for the red squares. These latter correspond 
to the brightest peaks in Fig. \ref{fig-bragg}.

\vspace{0.5cm}
\section{Conclusions}
\label{sec-concl}
 We develop a very straightforward method to calculate
the energy-band dispersion for quasicrystals as a function of the extended
Brillouin zones. Our strategy allows to do without the rough approximation 
exploited usually in the literature to assume that the first Brillouin 
zone can be considered as the fundamental one.
The connection between the spectrum and the quasi-momentum space raises
the natural connection between the integrated density-of-states per band
and the area of the corresponding Brillouin zone, transposing the gap
labelling to a Brillouin zone labelling.
Moreover we label the Brillouin zones by labelling the Bragg vectors.
By using the recursive law of generation of a Penrose-tiled quasicrystal 
we find an analytical expression that gives the Bragg vectors
and we are able to identify the ones that correspond to the brightest
Bragg peaks. 

Our study offers the realistic possibility to measure the gap labelling
within an optics experiment \cite{Man2005}, polaritons \cite{Bloch-fibo} 
or with ultracold atoms \cite{Greiner2001}.
In this context, the extended Brillouin zones can be measured one by one
({\it i}) transfering a Bose-Einstein condensate in the corresponding band
by a Raman transition,
({\it ii}) populating homogenously the band {\it via} Bloch oscillations,
({\it iii}) ramping down the potential to project the band over 
the free-particle
paraboloid and ({\it iv}) doing a time-of-flight imaging to have access to the
(quasi-) momentum distribution \cite{Greiner2001,Wirth2011,GreinerPhD}. 
The measure
of the areas of the different momentum distributions, that correspond
to different Brillouin zones, 
would provide an experimental proof of the gap labelling theory.

\ack We acknowledge Jose Aliste for useful discussions. 

  \def\newblock{\hskip .11em plus .33em minus .07em} 


\end{document}